\documentclass[namedreferences]{SolarPhysics}
\usepackage[optionalrh]{spr-sola-addons}
\usepackage{epsfig}
\usepackage{graphicx}
\usepackage{color}
\usepackage{rotating}
\usepackage{url}

\begin{document}
\begin{article}

\begin{opening}

\title{The G-O Rule and Waldmeier Effect in the Variations of the
 Numbers of Large and Small Sunspot Groups}

\author{J.\ JAVARAIAH}

\runningauthor{J. JAVARAIAH}

\runningtitle{VARITIONS IN THE NUMBERS OF  LARGE AND SMALL SUNSPOT GROUPS}

 \institute{Indian Institute of Astrophysics, Bangalore-560 034, India.\\
email: \url{jj@iiap.res.in}\\
}

\begin{abstract}
We have analysed the combined Greenwich and Solar Optical Observing 
Network (SOON)
 sunspot group data during the period of 1874\,--\,2011 and determined  
 variations in the annual numbers (counts) of the small
 (maximum area $A_{\rm M} < 100$ millionth of solar hemisphere, msh), 
 large ($100 \le A_{\rm M} < 300$ msh),  
and big ($A_{\rm M} \ge 300$ msh) spot groups. 
We  found that the amplitude 
of an even-numbered   cycle of  the number of large groups 
 is smaller  than that of  its immediately following odd-numbered
 cycle. This is consistent with  the well known 
Gnevyshev and Ohl rule or G-O rule
 of solar cycles, generally described  by using the Z\"urich
sunspot number ($R_{\rm Z}$).  During cycles 12\,--\,21 the G-O rule  
holds good for the variation in the number of  
small groups  also, but
  it is violated by   
 cycle pair (22, 23) as in the case of $R_{\rm Z}$.   This   
 behavior of the variations in the 
 small groups is largely  responsible  for the anomalous 
behavior of  $R_{\rm Z}$ in cycle pair (22, 23). 
 It is also found that   the amplitude of an  odd-numbered
 cycle of the number of small groups is larger   than that of  its immediately 
following even-numbered cycle. This can be called as `reverse G-O rule'. 
 In the case of the number of the big groups, both   cycle pairs 
 (12, 13) and (22, 23)   violated the G-O rule.
In many cycles the positions of the peaks 
of the small, large, and big groups   are different and   
 considerably differ with respect to the 
corresponding 
positions of the $R_{\rm Z}$ peaks.   
 In the case  of cycle~23, the corresponding cycles of the
 small and large   groups  are largely symmetric/less asymmetric
(Waldmeier effect is weak/absent) with their maxima taking place two years 
later than that of $R_{\rm Z}$. 
The corresponding 
  cycle of the big groups is more asymmetric (strong Waldmeier effect)
with its maximum epoch taking place at the  same time as  that of $R_{\rm Z}$.
\end{abstract}
\end{opening}

\section{Introduction}
Studies on variations in solar activity are important for understanding
 the mechanism behind the solar activity and solar cycle,  and also for 
predicting the level of activity~\cite{hat09,pet10}.
The properties of solar cycle are generally 
described by the Z\"urich or international sunspot number,   
 $R_{\rm Z} =  k (10 g + f)$, 
where $k$ is a correction 
factor for the observer, $g$ is the number of identified sunspot groups, 
and $f$ is the number of individual sunspots.
Several other solar activity 
indices  well correlate with $R_{\rm Z}$~\cite{hw04}.
 However, there are 
  noticeable differences in the epochs of the peaks of 
 $R_{\rm Z}$ and  other activity indices  in  some solar cycles  
($e.g.$, see~\inlinecite{rr08} and references therein). 
 It seems there are also considerable    
differences between  
the  epochs of the maxima of
 sunspot cycles and the corresponding  cyclic variations in 
the sunspot field
 strength ~\cite{pevt11}. 
 It is believed that the area of a 
 sunspot or a
 sunspot group  has a better physical significance   than $R_{\rm Z}$  
because the area is a better 
 measure (proxy) of  solar magnetic flux than $R_{\rm Z}$~\cite{dg06};
 an area of 130 msh (millionths of solar hemisphere; 1 msh 
$\approx 3\times 10^6$ km$^2$) corresponds approximately to 
 $10^{22}$ Mx (maxwell)~\cite{ws89}.
 There exists a high correlation between  solar cycle variations
 of $R_{\rm Z}$ and sunspot area~\cite{hwr02,hw04}.
 There are some 
 minor but noticeable  differences in the variations of  $R_{\rm Z}$  and 
 sunspot group area. 
\inlinecite{dgt08} 
 reported that the well-known Waldmeier effect  (inverse relationship between 
the rise time and the amplitude of a cycle) 
 does not exist in the case of sunspot area. 
The  epochs of the maxima of some cycles of sunspot number and sunspot area 
 are  different.
  For example, in case of  cycle~23, 
 the epoch of maximum of the sunspot number 
 was in 2000, whereas the maximum of the sunspot area  was in 2002~\cite{kbr10,jj12}.

The sunspot cycles are numbered in chronological order from the cycle 
that started 
from the year 1755, and are  known as Waldmeier cycle numbers. 
Many  characteristics of sunspot cycles are known 
(see \opencite{hw04}).
 Usually  the  time series  $R_{\rm Z}$  
 is used to reveal most of the characteristics of the solar cycle.
 However, Hoyt and Schatten (\citeyear{hk88a}, \citeyear{hk88b}) 
devised a number index based solely on the number of observed sunspot groups.
 The group sunspot number, $R_{\rm G}$, gives a more 
complete and longer data set than  $R_{\rm Z}$~\cite{hwr02,hw04},
 but the time series  $R_{\rm G}$ ended in 1995. 
 None of the characteristics of the solar cycle is fully understood 
so far. One of the most  prominent and  fundamental characteristics is 
    the differences in the 
amplitudes and the lengths of different solar cycles. According to 
the well-known Gnevyshev-Ohl rule~\cite{gh48} or G-O rule,
  the amplitude of an odd-numbered 
cycle is higher than that of the preceding even-numbered cycle. 
 However, there are 
 instances of violation of this rule, {\it viz}, cycle pairs (4, 5), (8, 9), 
and (22, 23). So far no plausible method is available to 
 predict the violation of the G-O rule (except that it may be possible from the
 epochs of 
 the retrograde motion of the Sun about the solar system 
 barycenter, as suggested by ~\inlinecite{jj05}).
 Using the data on sunspot groups during the period 1879\,--\,2004,
 \inlinecite{jbu05a} found that the solar equatorial rotation rate during
 an odd--numbered
 sunspot cycle   correlates well with  the equatorial rotation rate of the
 preceding even--numbered sunspot cycle, which is similar to the 
 G-O rule in sunspot activity. They also found that  the latitudinal gradient 
of  solar rotation during  an even--numbered cycle 
 correlates well with   that of the preceding odd--numbered cycles.
 These results seem
 to imply that the G-O rule is related to   the basic mechanism of solar 
activity and solar cycle.

 Typical sizes of sunspots range from
 10 
to $10^3$ msh.
Although single sunspots
are common,  the majority of sunspots belong to  sunspot groups.
 Sunspot groups are often large
 and complex.
 It is generally believed that large sunspot groups also  live long.
In fact, there is a rule of  proportionality between
the maximum area ($A_{\rm M}$) of a sunspot
group and its life time ($T$) (first noticed by~\inlinecite{gne38} and
formulated by~\inlinecite{wald55}; see also~\opencite{pvd97}):
 $\frac {A_{\rm M}} {T} \approx   10\ {\rm msh\ day}^{-1}$.
 However, the relationship between the area and the life time of
sunspot groups may be exponential rather than 
 linear~\cite{jj03a}. 
 Properties such as the  rotation rate, meridional motion, tilt angle, 
{\it etc.}, 
 of  sunspots and sunspot groups depend on their life time and size 
as well as their age 
  ($e.g.$,~\opencite{war65}, \citeyear{war66}; \opencite{hgg84}; 
\opencite{bal86}; \opencite{jg97}; \opencite{jj99}; \opencite{siva07}). 
 Studies 
on these properties of sunspot groups  may  
provide information on the subsurface dynamics 
 of the Sun ($e.g.$, \opencite{how96}; \opencite{jg97}; \opencite{kmh02}; 
\opencite{siva03}, 2007, 2010). 
Therefore, the studies on the variations in the 
 numbers (counts)  of  
 sunspots and sunspot groups in different sizes  
 look to be important for understanding the basic mechanism of  
solar activity and 
solar cycle, and also the relationship between   sunspots and other
 activity indices
 ($e.g.$, \opencite{kilc11}; \opencite{cl12}).   
In the present paper we have analyzed the sunspot group data
 during 
the period 1874\,--\,2011  and studied the variations in the annual numbers 
 of sunspot 
groups of different sizes. 
Particularly  we have concentrated on  
 the G-O rule and the Waldmeier effect in
 the variations of small and large sunspot groups, and
 their implications.

In the next section we will describe
  the data and the method of analysis. In Section~3 we will
 describe the results
 and in Section~4 we will  present conclusions and a brief discussion.

\section{Data and Analysis} 

Here we have used Greenwich and Solar Optical Observing Network (SOON)
sunspot group data during the period of  May 1874 to May 2011
(taken from \break {\tt http://solarscience.msfc.nasa.gov/greenwch.shtml}).
 These data
include the observation time (the Greenwich data contain the date with the
fraction of a day, in  the SOON data
 the fraction  is rounded to 0.5 day),
heliographic latitude ($\phi$) and
longitude ($L$), central meridian distance (CMD), and
corrected umbra and whole-spot areas (in msh), {\it etc.}, of sunspot
groups for each day of observation.
The
positions of the groups  are geometrical  positions of the
centers of the groups.
The Greenwich data (May 1874 to December 1976) have been compiled
 from the majority of the white
light photographs which were secured at the Royal Greenwich Observatory
and at the Royal Observatory, Cape of Good Hope. The gaps in their
observations were filled with photographs from  other observatories,
 including the Kodaikanal Observatory, India.
The SOON data (January 1977 to May 2011)  include measurements made   by the
United States Air Force (USAF) based on 
 sunspot drawings obtained by a network of the observatories
 in Boulder,  Hawaii, and so on.
David Hathaway  scrutinized
the Greenwich and SOON data and produced a reliable
continuous data series
from 1874 up to date.
 In case of SOON data, we increased area by a factor of 1.4.
 David Hathaway found this correction was necessary to
 have a combined homogeneous
 Greenwich and SOON
data (see the aforementioned web-site of David Hathaway).
The combined  Greenwich and SOON sunspot group data  are the largest
 available, 
 reliable data that include  
 the  positions and areas of sunspot groups. 

  If $A_1$, $A_2$,...,$A_n$ denote the areas (corrected for the 
foreshortening effect) of all the sunspots in  a
 sunspot group  observed at times 
$t_1$, $t_2$,...,$t_n$  during the life time of the 
sunspot group  $T = t_n - t_1$ (days), then  the  
maximum area  is defined as $A_{\rm M} = {\rm max} (A_1, A_2,...,A_n$),
 where $n = 2, 3, \dots$.  
We have used here only the sunspot groups which had $T \ge 2$ days.   
On the basis of  $A_{\rm M}$ values   we have classified sunspot groups 
 into three classes  as follows:
 small sunspot groups (SSGs: $A_{\rm M} < 100$ msh),
large sunspot groups (LSGs: $100 \le A_{\rm M} < 300$ msh), 
and big sunspot groups (BSGs: $A_{\rm M} \ge 300$ msh).
We used the data on only those sunspot groups whose birth and death occurred 
within a disc passage. 
 That is, we have not used the  sunspot groups whose central meridian distance 
 $|{\rm CMD}| > 75^\circ$ in any day of their respective life times. This reduces 
the foreshortening effect and helps to obtain the maximum area of a  sunspot
 group unambiguously. 
Each appearance of a recurrent group is treated 
as an independent group. 
Thus, $T \le 12$ days.
We determined the numbers (counts) NSGs, NLGs, and NBGs of SSGs, LSGs 
and BSGs, respectively, 
 for each year during the period 1874\,--\,2011.

\section{Results}

Figure~1 shows  the variations  in the annual NSGs, NLGs, and NBGs 
 during the period 1874\,--\,2011. In the same figure we have also shown 
the variation of  annual  $R_{\rm Z}$    
 taken from the website, 
{\tt ftp://ftp.ngdc.noaa.gov/STP\break /SOLAR\_DATA/SUNSPOT\_NUMBERS/INTERNATIONAL/yearly/YEAR.PLT}).
As can be seen in this figure, 
 each of these parameters   shows the  
11-year period solar cycle variation. Further,  there is an indication that 
as the size of 
the group decreases its counts  increase. That is,  in most of the
 time the curve of NLGs is 
 above the curve of NBGs and the curve of NSGs is above that of NLGs.
This is consistent with the well-known result that the smallest regions 
dominate the global flux emergence rate (\opencite{zirin87}).  
  At the epoch of the maximum  of cycle~21 the NSGs is largest, $i.e.$, there 
the NSGs is larger than   even the NSGs is  at the epoch of the maximum
of the largest solar cycle, 19.    
Recently, \inlinecite{kilc11}  analyzed the Rome Observatory sunspot group data
 for solar cycles 20 and 21 and Learmonth Solar Observatory data for  cycles 
22 and 23. By using the   Zurich classification of the sunspot groups, 
these authors   found that in cycle~23 the number of large sunspot groups 
 is higher 
 when compared to those for cycle~22.
A similar tendency  can be seen in 
Figure~1, {\it i.e.}, the peak in NLGs  of cycle~22 is smaller than 
  the corresponding peak of cycle~23 (the peak in NBGs of cycle~22 is only 
slightly larger than that of cycle~23). Hence,
 the aforementioned result that was found  by~\inlinecite{kilc11}  
 is confirmed here.  
 In fact, in this figure one can see that  the amplitude 
of an even--numbered   cycle seen in NLGs  
 is smaller  than that of  its immediately following odd--numbered cycle.
 This relationship of  even- and odd--numbered cycles   is
 consistent with   
    the well-known G-O rule 
of modulation in the amplitudes  of solar cycles. 
In the case of $R_{\rm Z}$  cycle pair (22, 23) violated 
 this rule, but in the case of  NLGs  
 cycle pair (22, 23)  satisfied it.    
The behavior of  NSGs  is similar to  the  behavior of  
$R_{\rm Z}$. However, in the case of the former we can see another 
striking systematic behavior that did not exist in the case of the latter. 
That is,  the amplitude of an  odd--numbered
 cycle in NSGs is larger than that of  its immediately 
following even--numbered cycle,  a behavior that is opposite 
  to the G-O rule. 
 This `reverse G-O rule' 
  indicates that in the case of NSGs  the amplitude 
of the current cycle~24 will be smaller than the previous 
cycle 23. In the case of NBGs the G-O rule
 is violated by both
 cycle pairs (12, 13) and (22, 23) [the amplitude of cycle~22 is  only
 slightly larger (almost equal) than  that of cycle~23], but this
 is not reflected
in $R_{\rm Z}$  because in a given time interval  NBGs is considerably
 smaller than  NSGs ($R_{\rm Z}$ gives an equal weight to all sunspots 
and to all sunspot groups ($e.g.$, \opencite{cl12})).
 The same is true in the case of NLGs during cycles
22 and 23, where $R_{\rm Z}$ violated the G-O rule and NLGs satisfied it.  
   The `reveres G-O rule'  does not exist in the amplitude modulations of
  NLGs and NBGs cycles. However, because of possibly insufficient size of 
  data,  it may be cautioned that 
 all the results above  are only suggestive rather than compelling.

\begin{figure}
\centerline{\includegraphics[width=\textwidth]{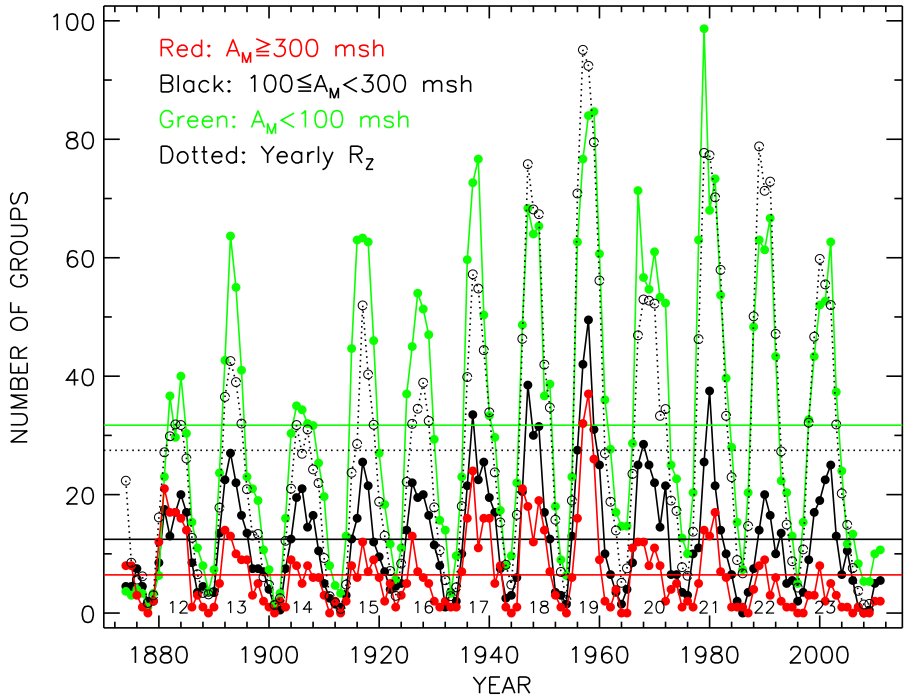}}
\caption{Variations in  the annual numbers of the  small (green curve), 
large (black curve), and big spot groups (red curve),  divided by 3, 2, and 1 
respectively, during the period 1874\,--\,2011 (Note: the data in 1874 and
 2011 are incomplete). The open circles connected by the dotted lines
 represents the values   of 
$R_{\rm Z}$ divided by 2. 
 The horizontal lines represent the respective mean values. Near the 
 maximum of each cycle the corresponding Waldmeier cycle number 
is indicated.}  
\label{fig1}
\end{figure}

In Figures~2, 3, and 4  we have compared the variations in the annual 
NSGs, NLGs, and NBGs, respectively, in different solar cycles. 
Table~1 shows the positions of the peaks of NSGs, NLGs, and NBGs 
with respect to 
 the positions of the corresponding $R_{\rm Z}$
 peaks of  cycles 12\,--\,23, obtained from  Figures~2\,--\,4. 
 In this table it  can be seen that in many cycles the 
positions of the peaks of 
NSGs, NLGs, and NBGs  are different and they also considerably 
deviate from  the 
corresponding 
positions of the $R_{\rm Z}$ peaks. 
 In the case of the largest solar cycle~19, the positions of 
the peaks of all the three classes of the 
 sunspot groups are the same, $i.e.$, they  
occurred at  one year later
than the peak of $R_{\rm Z}$  (a similar behavior is  also 
seen in the amounts (annual rates) of the growth and decay of sunspot 
groups~(\opencite{jj12})).
 A negative  value of the position of the peak of a given class of groups 
 implies 
that the 11-year cycle of this class of groups is more asymmetric  
(strong Waldmeier effect) than the corresponding cycle of $R_{\rm Z}$. 
 In the case of cycle~23, the corresponding cycles in  
 NSGs and NLGs were largely symmetric (or less asymmetric), namely the 
Waldmeier effect was weak or absent, and  the corresponding cycle in 
 NBGs  was more asymmetric (strong Waldmeier effect).  
The extended declining phase ($i.e.$, beyond the length  
of a normal cycle)  of this  long cycle~23  
included only small sunspot groups. The occurrence of big groups stopped 
much earlier, in 2004. 
 \inlinecite{kilc11} found  that the numbers of  small and large sunspot 
groups show similar time variations during cycle 22, and  cycles 20, 21, 
and 23 show different behavior. Namely, the peak of 
NSGs was 
 during the first maximum of $R_{\rm Z}$ and that NLGS 
  was at the 
second maximum of $R_{\rm Z}$. 
In Table~1  it can be seen that this result is  applicable only to 
    cycle~21. It may be    
 suggested that   small sunspot groups
substantially contributed to the 
second peak in $R_{\rm Z}$  of cycle~23. Thus,  the aforementioned  result in 
 \inlinecite{kilc11} is not 
 confirmed in the present analysis. 

\begin{figure}
\centerline{\includegraphics[width=\textwidth]{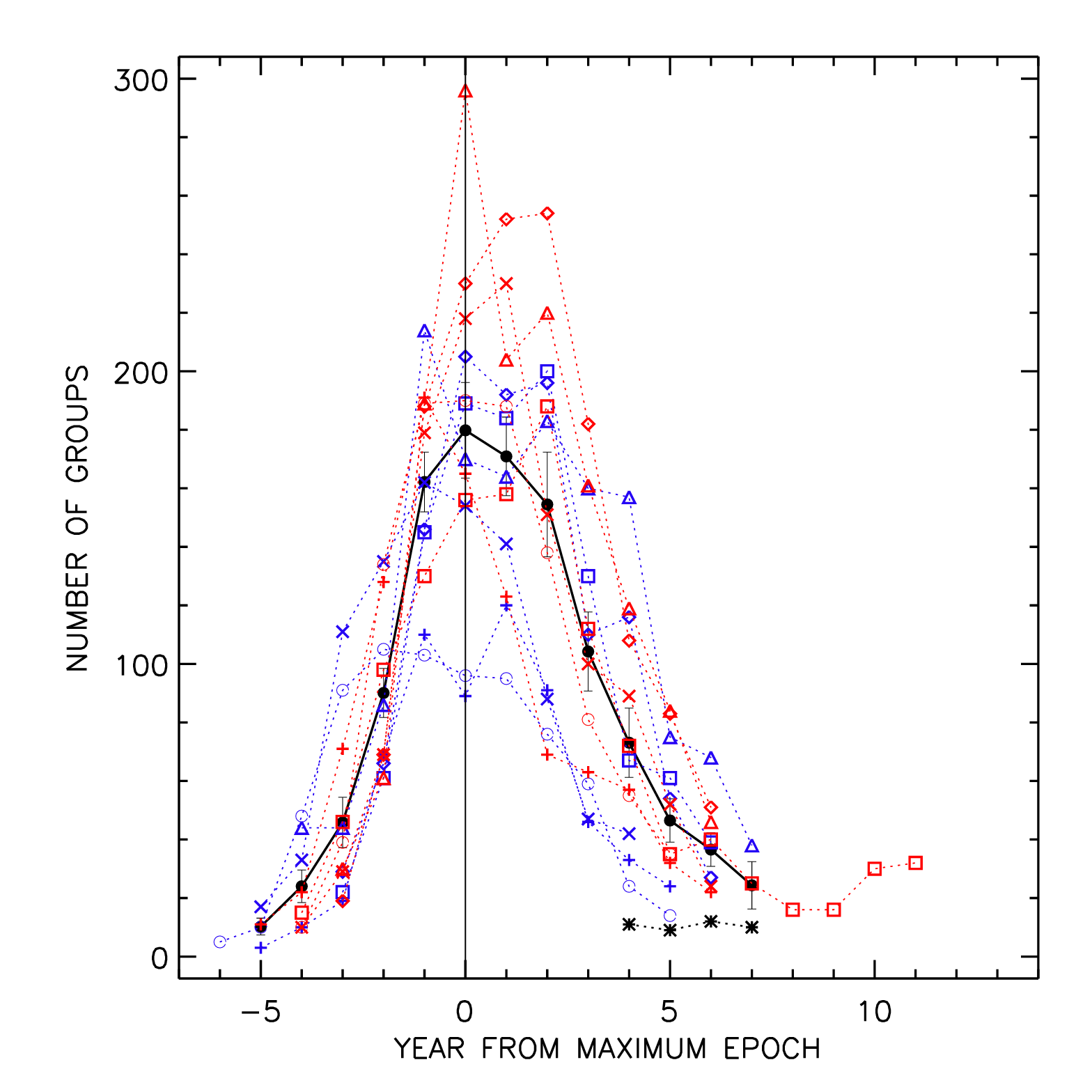}}
\caption{The annual number of 
  small spot groups (NSGs)
{\it versus} the year from the maximum epoch of the solar cycle.
The blue and red colors are used for even- and odd-numbered
cycles, respectively. (For the sake of clarity  black color 
 is used for cycle~11.).  Different symbols are used for
different cycles (numbers are given in the parentheses): asterisks (11),
pluses (12 and 13),
open-circles (14 and 15), crosses (16 and 17), diamonds (18 and 19),
triangles (20 and 21), and squares (22 and 23).
 The filled circles connected by the solid lines represent
 the mean solar cycle variation determined from the  values of annual numbers.
 The error bars represent  the standard error 
(standard deviation is divided by the square-root of the number of 
data points minus one).
There is only one
data point  at  years -6 (beginning of cycle~14), 8 (end of cycle~23) and
9-11 (first three years of cycle 24).}
\label{fig2}
\end{figure}

\begin{figure}
\centerline{\includegraphics[width=\textwidth]{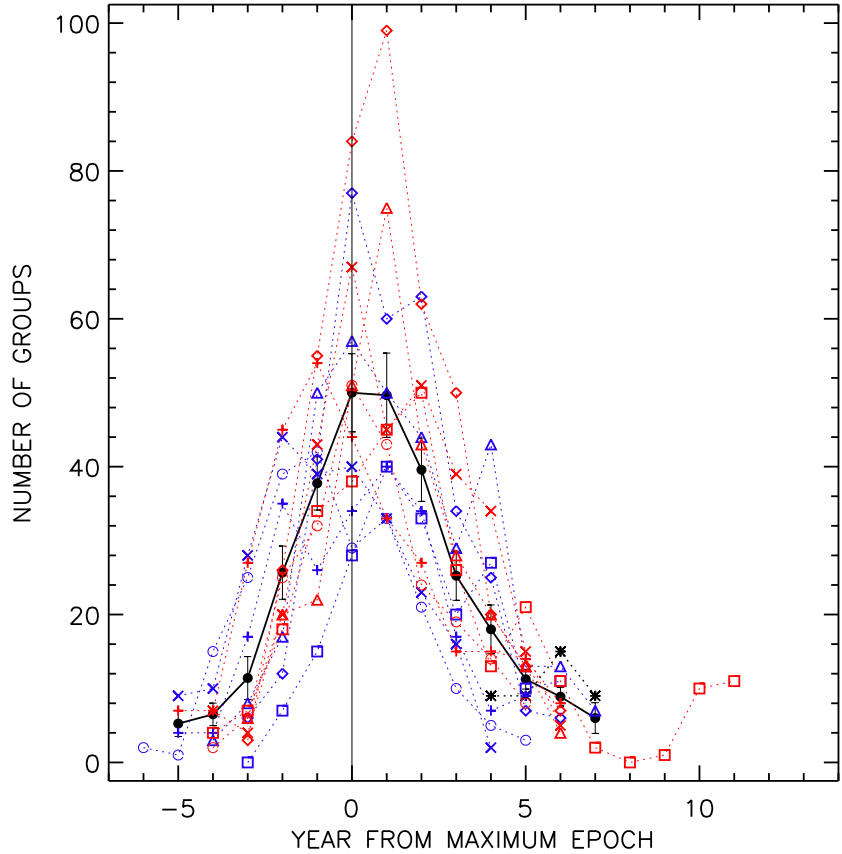}}
\caption{The same as Figure~2 but for the number of large sunspot groups (NLGs).}
\label{fig3}
\end{figure}

\begin{figure}
\centerline{\includegraphics[width=\textwidth]{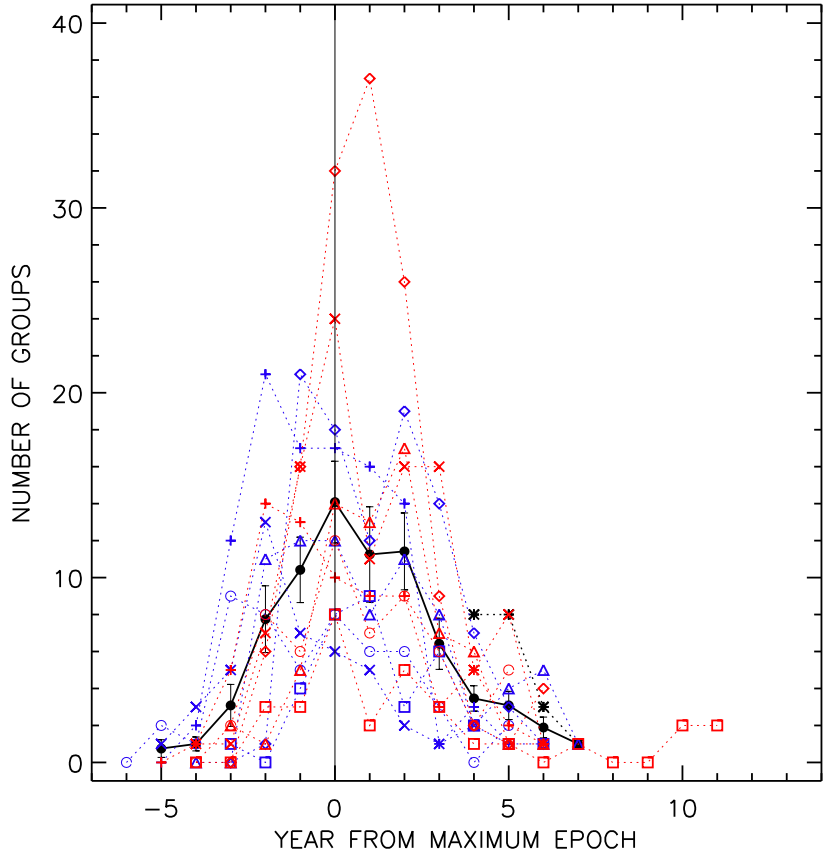}}
\caption{The same as Figure~2 but for the number of big sunspot groups (NBGs).}
\label{fig4}
\end{figure}

\begin{table}
\flushleft
   \caption{Positions of the peaks of NSGs, NLGs, and NBGs from 
the corresponding epochs ($Y_{\rm M}$) of the maxima of solar cycles 
 12\,--\,23, as can be seen in Figures~2, 3 and 4.}

\begin{tabular}{lccccc}
\hline
  \noalign{\smallskip}
Cycle & $Y_{\rm M}$& \multicolumn{3}{c}{The peak positions (in year) of} \\
  \noalign{\smallskip}

&&NSGs&NLGs&NBGs\\
\hline
  \noalign{\smallskip}

12 (1878 - 1888) &1883    & $ +1$  &  $ +1$  &   $-2$ \\
13 (1889 - 1900) &1894     & $-1$  &  $ -1 $ &   $-2$\\
14 (1901 - 1912) &1907      &$-2$(flat)  &  $-1$  &   $-3$\\
15 (1913 - 1922) &1917     & $-1/0$ &  $ 0$  &    $0$\\
16 (1923 - 1932) &1928    &  $-1/0$  & $-2$ &    $-2$\\
17 (1933 - 1943) &1937    & $0/+1$  &  $0$ &     $0$\\
18 (1944 - 1953) &1947  &   $ 0 $  &   $ 0$ &   $ -1$\\
19 (1954 - 1963) &1957   &  $ +1/+2$ & $+1$  &  $ +1$\\
20 (1964 - 1975) &1968   &  $ -1 $ &   $ 0$  &  $ -1/0$\\
21 (1976 - 1985) &1979   &  $  0 $&   $ +1$ &   $ +2$\\
22 (1986 - 1995) &1989   &  $ +2 $ &   $ +1$ &  $  +1$\\
23 (1996 - 2008) &2000   &  $ +2 $ &   $ +2$ &  $  0$\\ 

\hline
  \noalign{\smallskip}

\end{tabular}

\end{table}

Figure~5 shows the cycle-to-cycle variations in the   
 annual NSGs, NLGs, and NBGs averaged over each cycle.
In the same figure we have also shown 
 the cycle-to-cycle variation in the amplitude  ($R_{\rm M}$,  
 the  largest  value of smoothed monthly 
mean sunspot numbers in each cycle). The values of $R_{\rm M}$ are taken from the website, 
{\tt ftp://ftp.ngdc.noaa.gov/STP/SOLAR\_DATA/SUNSPOT\_NUMBERS}.
As can be seen in this figure, the patterns of the long-term variations 
 in the average annual 
 NSGs, NLGs, and
 NBGs are  similar to that of $R_{\rm Z}$. 
 The long-term trend indicates that these quantities, particularly 
NSGs, continue to decline for a few more cycles. 
 It may be  suggested that 
the  long-term variation shows the  largest amplitude  in 
the case of  NSGs and its period is also  long
 (180 years) in this case. 
The NBGs seems to have a 90-year cycle with minimum  at cycle~16 and the 
maxima at cycles~12 and 19. 
There was a downward trend in NSGs from cycle~21. This   
 indicates that NSGs would be small in cycle~24 and even in cycle~25. 
Since in any cycle (or its sub-interval) the SSGs are  
majority,  we can suggest that the amplitude of  cycle~24 in $R_{\rm Z}$
will be smaller than that of cycle~23. This is consistent with a low amplitude 
predicted for cycle~24~\cite{jj08,kbr12}.

\begin{figure}
\centerline{\includegraphics[width=\textwidth]{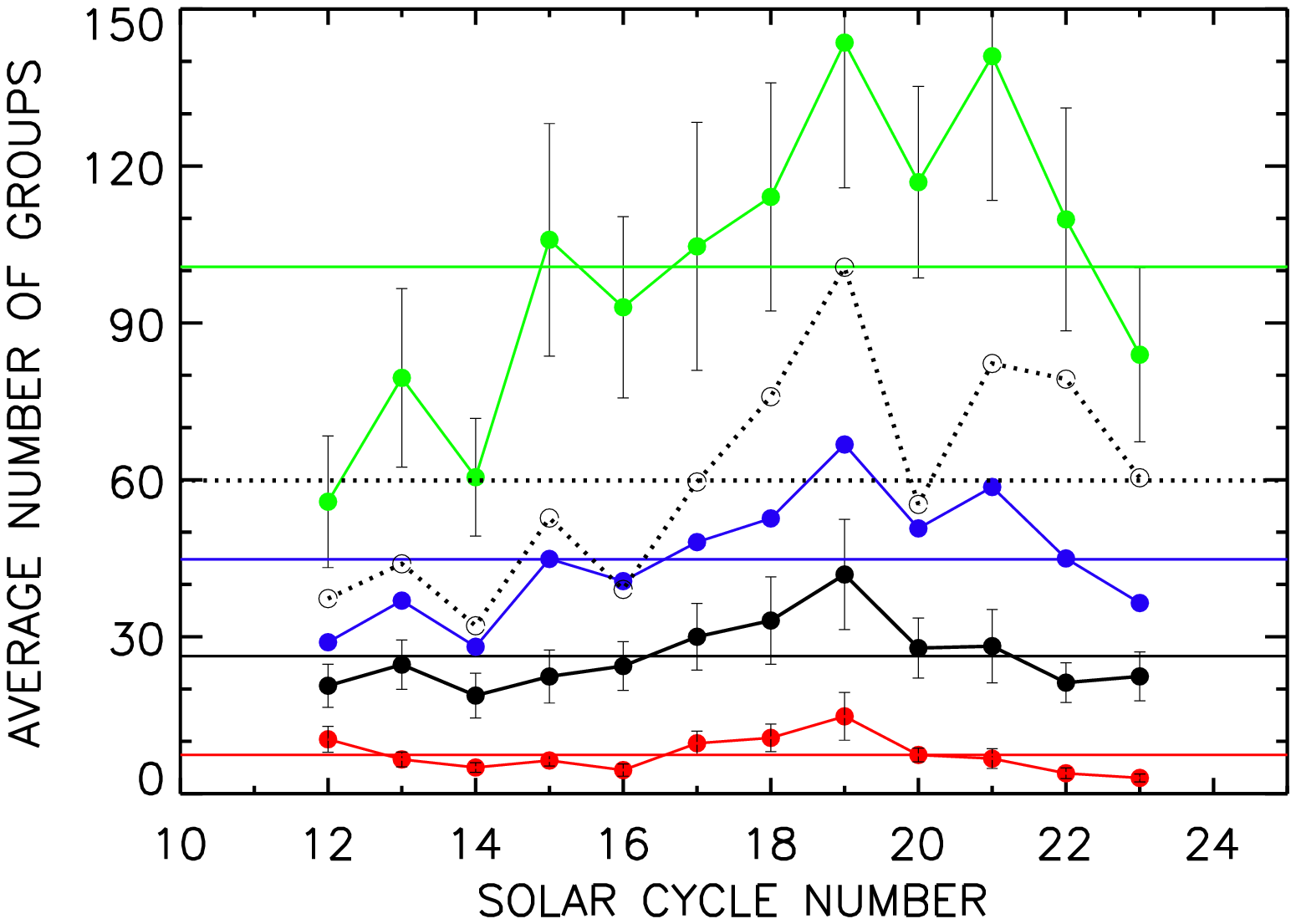}}
\caption{Cycle-to-cycle variation in the annual numbers of  
 small (green curve), large (black curve), and big (red curve) 
sunspot groups averaged over each cycle. 
The blue curve represents the corresponding  mean of all these three classes.  
The open circles connected by the dotted lines represent the 
 variation in   $R_{\rm M}$/2 ($R_{\rm M}$ represents the maximum amplitude of a cycle, 
$i.e.$,  the largest value of smoothed monthly 
mean sunspot numbers which are taken from the website, 
{\tt ftp://ftp.ngdc.noaa.gov/STP/SOLAR\_DATA/SUNSPOT\_NUMBERS}).
 The horizontal lines represent the 
respective mean values.} 
\label{fig5}
\end{figure}

\section{Conclusions and Discussion}
From the above analyses of  a large data set on sunspot groups the following 
 conclusions can be drawn: 
\begin{enumerate}
\item  The amplitude 
of an even--numbered   cycle of  NLGs
 is smaller  than that of  its immediately following odd--numbered cycle.
 This is consistent with  the well-known 
 G-O rule of solar cycles. Obviously,  the amplitude of cycle~22 of 
NLGs is smaller than that of cycle~23. This is in  line with the 
conclusion drawn by~\inlinecite{kilc11}.  
\item NSGs  also satisfies the even--odd cycle rule, but  
 cycle pair (22, 23) violated the G-O rule, {\it i.e.}, in this cycle pair 
 the behavior of NSGs 
 is  similar to that of   $R_{\rm Z}$.
 It is also found that   the amplitude of an  odd--numbered
 cycle of NSGs  is larger   than that of  its immediately 
following even--numbered cycle. This can be called as a `reverse G-O 
 rule'.
\item In the case of NBGs,   cycle pairs 
 (12, 13) and (22, 23) show similar behavior, {\it i.e.}, both  
 violated the even-odd cycle rule. The violation of the 
G-O rule in NBGs   is not reflected in 
$R_{\rm Z}$  because contributions from   NBGs to $R_{\rm Z}$ are 
 relatively small.    
\item In many cycles the positions of the peaks 
of NSGs, NLGs, and NBGs  are different, and they also deviate     
 considerably from the   
corresponding peak  
positions of  $R_{\rm Z}$. 
 In the case of cycle~23, 
the maxima of NSGs and NLGs  are at 2002, whereas the maximum of NBGs 
 is at 2000, $i.e.$, at the  same epoch of  
the maximum of $R_{\rm Z}$. The corresponding  cycles in 
 NSGs and NLGs  are largely symmetric or less asymmetric
(the Waldmeier effect is weak or absent), and  the corresponding 
cycle in NBGs  is more asymmetric (strong Waldmeier effect). 
\item The amplitude of the long-term variation
 is large in the case of NSGs,  and its  period  ($\approx$ 180 years)
 is also long in this case. An approximate 90-year cycle is seen in NBGs. 
The long-term trend in the number of small groups  implies that the current 
cycle~24 is weak. 
\end{enumerate}

The studies of rotation rates of sunspot  groups
of different life times and sizes indicated that 
the magnetic structures of small and large/big  groups anchor near the surface 
(near 0.95$R_\odot$, here $R_\odot$ 
is the radius of the Sun)
and relatively deeper layers  (even reach near 
to the base of the convection zone) of the 
Sun's convection zone, 
respectively~(\opencite{jg97}; \opencite{kmh02}; \opencite{siva03}, 2004).
  Such a study also suggested that small sunspot groups may be 
the fragmented or the branched parts of the 
large/big sunspot groups~\cite{jj03a}. In other words,  large/big 
sunspot groups 
may be the products of a deep dynamo mechanism, whereas  small 
 sunspot groups 
may be the product of a surface dynamo mechanism.
Thus,  our conclusions 1 and 
2 above  imply that the G-O rule represents a deep rooted global property of 
the solar cycle. Because of their large numbers the small sunspot 
groups 
 mainly contributed to the behavior of cycle pair (22, 23) to 
violate the G-O rule. That is, the violation of the G-O rule in $R_{\rm Z}$
 may be largely 
related to the surface (local) dynamo,  for  which 
convection and  meridional flows may  have major roles 
(cancellation of magnetic flux). The `reverse G-O rule' 
found in the variation of the number of small groups (conclusion 2 above)
 is also related to the surface (local) dynamo mechanism.  
  
The existence of approximate 90-year and 180-year periodicities  has been
  known 
in sunspot activity. In the present analysis they are seen (visualized) 
 in the variations of the numbers of large/big sunspot groups 
and of small sunspot groups, respectively 
(conclusion 5 above), suggesting that 
the former and latter are  related to the dynamics of the deeper  and the   
near-surface layers, respectively.  

A `reverse G-O rule'~\cite{jbu05a}  and  a 90-year 
periodicity~\cite{jj03b,jbu05b}
 were also found in the latitude gradient of solar rotation 
determined from the sunspot group data. 
 Conclusions~2 and 5 above 
 indicate that  rotations of  small and  big  sunspot groups largely 
contribute to  the reverse G-O rule and the 90-year periodicity of 
the latitude gradient, respectively. 
 Conclusion~2 above also     
  indicates  that in terms of  the number of SSGs   
the amplitude 
of the current cycle~24  
 will be  smaller  than  
that of the previous cycle~23.

The maximum in $R_{\rm Z}$ of  solar cycle
 is not smooth sometimes. Two or more peaks can be identified
during the solar maxima and are called Gnevyshev peaks, because 
 this splitting of activity was
identified for the first time by 
Gnevyshev~(\citeyear{gne67}, \citeyear{gne77}). 
The time interval between these peaks, where the level of activity is 
relatively low, 
 is known as the Gnevyshev gap (see the review by~\opencite{sto03}).
\inlinecite{gopal03} found that during cycle~23 
the  occurrence rate of coronal mass ejections (CMEs) 
 peaked two years after 
the maximum epoch (2000) of this cycle.
\inlinecite{det04} reported that not only  
  $R_{\rm Z}$ but also all other  activity  indices, 
 including facular area, F10.7 flux, {\it etc.},  
show a double peak structure near the maximum of solar cycle 23, and 
 these  have the highest peak
 in the year 2002. 
\inlinecite{kbr10} showed that the occurrence peak of CMEs 
is close to the peak of the 
 sunspot  group 
area.  More powerful flares seem   to occur after the maximum 
epoch of $R_{\rm Z}$ and the maximum of the annual number of X-class flares
 took  place  close to
 2002~\cite{tan11}. \inlinecite{jj12} found that the amounts (annual rates)  of 
growth and decay 
of  magnetic flux  in sunspot groups  in a given time
 interval (year)
  correlate well with  the amount of 
magnetic flux available in that interval. Hence, they have claimed that 
 the solar cycle variation in the decay of sunspot groups 
has a substantial 
contribution to the solar cycle variations in the 
solar energetic phenomena and the total solar irradiance (TSI).   
\inlinecite{kilc11} suggested  that the excess of 
large sunspot groups  during the declining 
phase of cycle~23 is responsible for the occurrence rate of CMEs and  
other activity indices 
to reach their maxima in the year 2002. 
 Our results  (conclusion 2 above)  
 indicate that there is  a considerable contribution from
the small sunspot groups concerning  
 the two--year delay in the maximum of CME occurrence rate 
  and other activity indices/energetic phenomena. 
That is,  
as suggested in   \inlinecite{jj12},    
 TSI and   solar 
 energetic phenomena such as  
 flares and CMEs seem to be largely related to the more evolved flux of 
sunspot groups.

\acknowledgements{The author thanks Dr. K. B. Ramesh for reading the 
 manuscript and for discussion/suggestions, and the  Editor in Chief Professor 
Takashi Sakurai for useful comments and suggestions.}

{}

\end{article}           
\end{document}